\documentclass[preprint,aps,amsmath,pre,showpacs]{revtex4-1}
\usepackage[utf8]{inputenc} 
\usepackage{graphicx}
\usepackage{amsmath,amssymb,amsthm,amsfonts}
\usepackage{epsfig}
\usepackage{xargs}
\usepackage{setspace}

\usepackage{soul} % Highlight using \hl{}
\usepackage{xcolor}
\definecolor{crimson}{HTML}{DC143C}
\definecolor{lightpink}{HTML}{FFB6C1}
\definecolor{lightskyblue}{HTML}{87CEFA}

\newcommand{\av}[1]{\left\langle#1\right\rangle}
\newcommand{\lrp}[1]{\left(#1\right)}
\newcommand{\lrc}[1]{\left[#1\right]}

\begin{document}

\title{Opinion Dynamics on Networks under Correlated Disordered External Perturbations}

\author{Marlon Ramos$^{1}$}
\author{Marcus A.M. de Aguiar$^{1,2}$}, 
\email{corresponding author:aguiar@ifi.unicamp.br}
\author{Dan Braha$^{2,3}$}

\affiliation{
$^{1}$Universidade Estadual de Campinas, Campinas, SP, Brazil\\
$^{2}$New England Complex Systems Institute, Cambridge, MA, United States\\
$^{3}$University of Massachusetts, Dartmouth, MA, United States
}

\begin{abstract}

We study an influence network of voters subjected to correlated disordered external perturbations,
and solve the dynamical equations exactly for fully connected networks. The model has a critical
phase transition between disordered unimodal and ordered bimodal distribution states, characterized
by an increase in the vote-share variability of the equilibrium distributions. The random
heterogeneities in the external perturbations are shown to affect the critical behavior of the
network relative to networks without disorder. The size of the shift in the critical behavior
essentially depends on the total fluctuation of the external influence disorder. Furthermore, the
external perturbation disorder also has the surprising effect of amplifying the expected support of
an already biased opinion. We show analytically that the vote-share variability is directly related
to the external influence fluctuations. We extend our analysis by considering a fat-tailed
multivariate lognormal disorder, and present numerical simulations that confirm our analytical
results. Simulations for other network topologies demonstrate the generalizability of our findings.
Understanding the dynamic response of complex systems to disordered external perturbations could
account for a wide variety of networked systems, from social networks and financial markets to
amorphous magnetic spins and population genetics.

\end{abstract}

%89.75.Kd Complex systems
%89.65.-s Social systems

\maketitle

\section{Introduction}
\label{intro}

Models of opinion formation, which explore the dynamics of competing opinions taking into account
the interactions among agents, have been extensively studied 
\cite{Nowak1990a, Galam2005a, sanmiguel05, chinellato2007a, castellano2009a,costa2011a, galam2012sociophysics, helbing2015saving, gonccalves2015social, chinellato2015dynamical, harmon2015anticipating, ramos2015does, braha2017voting}.
In their most basic form, these
models consist of voters, represented by nodes on a social network, having only two possible
opinions, 0 or 1. Each voter may change her mind by using various interaction mechanisms, for
example, randomly adopting the opinion of a connected neighbor (essentially a majority-vote rule,
see \cite{castellano2009a, liggett2012interacting, fernandez2014voter}), or by applying local majority rules \cite{Galam2005a, castellano2009a, galam2012sociophysics}. The stochastic dynamics of these
simple interaction models ultimately leads to a uniform state corresponding to the all-nodes-0 or
all-nodes-1 states where all voters share the same opinion. Obviously, consensus states are not
commonly observed in real-world applications. Accordingly, more realistic models of opinion dynamics
have been proposed that incorporate, among other features, social impact theory \cite{Nowak1990a,latane1981psychology,holyst2001social}, opinion
leaders and zealots \cite{chinellato2007a, chinellato2015dynamical, harmon2015anticipating, holyst2001social, mobilia2003does, galam2007role, acemouglu2013opinion, yildiz2013discrete, wu2012opinion, galam2016stubbornness, xie2011social, xie2012evolution, singh2012accelerating, mobilia2007role}, external influences and fields \cite{sanmiguel05, chinellato2007a, chinellato2015dynamical, galam1997rational, carletti2006make, kuperman2002stochastic, tessone2005system, gonzalez2005nonequilibrium, shibanai2001effects, mazzitello2007effects}, individual's biases \cite{galam2005heterogeneous, galam2017trump}, contrarians \cite{galam2004contrarian}, individual’s own current opinion \cite{shao2009dynamic, li2013non, calvao2016role}, word-of-mouth spreading
\cite{fortunato2007scaling}, non-overlapping cliques \cite{palombi2014stochastic}, or noisy diffusive process \cite{fernandez2014voter}.

Here we focus on an opinion formation model, which considers not only the role of internal
self-reinforcing influences between connected nodes in the network, but also the role of external
influences in opinion formation. More generally, these external influences represent the dynamic
response of a complex system to an external environment. We have previously modeled such external
influences as perturbations or modulations acting on all agents in the system \cite{chinellato2007a, chinellato2015dynamical, galam1997rational}, and have
obtained complete and exact results for fully connected networks and arbitrary, but constant
perturbations. Understanding the dynamic response of complex systems to external perturbations could
account for a wide variety of networked systems, from social networks and financial markets to
amorphous magnetic spins and population genetics. For example, in political elections with two
candidates \cite{braha2017voting} the external influence on uncommitted voters could represent numerous sources, which
convey consistent partisan bias in favor of one of the candidates over another, such as opinion
leaders or the mass media. In population genetics, the external influence can represent mutational
bias or selection towards one of two alleles of a gene in an evolving population of sexually
reproducing haploid organisms \cite{de2011moran}. In Ising-type spin models on crystalline 3-D lattices or
amorphous spin- glasses, the external influences correspond to temperature and external magnetic
field \cite{chinellato2015dynamical}. Finally, the model can also represent stock price movements (``up'' or ``down'') in a
network of stocks where the external influences correspond to ``news'' or new information that changes
perceptions of fundamental stock values \cite{harmon2015anticipating}.

Previously, we solved the dynamical equations of this model exactly for fully connected networks
under fixed external perturbations \cite{chinellato2007a, chinellato2015dynamical}, obtaining among others the equilibrium distribution of
voters’ opinions. We found a nontrivial dynamic behavior that can be divided into two regimes for
small and large external perturbations, displaying a disorder-to-order critical phase transition.
The disordered regime is characterized by skewed unimodal distributions with a peak corresponding to
the fraction of voters in the network that voted for opinion 1. The bistable ordered regime is
characterized by bimodal distributions in which two symmetry breaking phases may exist, similar to
the magnetization state in the Ising model below the critical temperature. The critical value of
this model, which marks the transition between disordered and ordered states, is a unique state with
a flat distribution of voters’ opinions. Under certain conditions, the spontaneous emergence of the
disorder-to-order phase switching is associated with an increase in the variability of the
equilibrium distribution of voters’ opinions.

The voter model mentioned above \cite{chinellato2007a, chinellato2015dynamical} is homogenous; that is, the strength of the external
influence was assumed to be the same for all voters in the network, rendering all voters identical.
This is obviously a limitation. Real complex systems inevitably contain random inhomogeneities,
which tend to disorder the system. In disordered systems, agents have individual attributes, which
are qualitatively the same for all of them, but differ quantitatively from one another—a
characteristic that is particularly an essential part of the statistical physics of social dynamics.

Here, we consider a disordered opinion dynamics, taking the aforementioned voter model under
external perturbations \cite{chinellato2007a, chinellato2015dynamical, harmon2015anticipating, braha2017voting} as our basic dynamical system. The disorder in our opinion
dynamics model arises from the randomness in the strength of the external perturbations. More
specifically, we assume that the parameters controlling the external influences are drawn from a
given probability distribution for each voter, and whose values change in time on a time scale
comparable to the voter opinion fluctuations—an example of annealed disorder. This is in contrast to
quenched disorder, where the values of the random variables vary from one voter to another according
to the external influence random distribution, but remain constant in time. This latter case can be
used as a proxy for disordered opinion dynamics in which the characteristic time scale of changes in
the external influence is much longer than the characteristic time scale for voter opinion
fluctuations. Thus, the disorder in the voter model depends crucially on the relative time scales of
the external influence and voter opinion fluctuations. If these two time scales are comparable with
one another -- as observed in real world applications of the voter model \cite{harmon2015anticipating, braha2017voting} -- then the disorder
should be considered to be annealed. We solve the dynamical equations exactly for fully connected
networks but also consider a quenched disorder that arises from randomness in the topology of the
network (i.e., the connectivity may be different for different voters). For quenched disordered
networks, the randomness in the network structure is fixed in each realization of the opinion
dynamics, and therefore the network does not evolve in time. This assumption is valid if the
characteristic time scale for changes in the network is much longer than the characteristic time
scale of the voter opinion dynamics. The voter model with quenched disordered networks is
analytically intractable, and is studied here using simulations.

In this paper, we address the question of how the equilibrium behavior (including its behavior near
the phase transition) of an influence network of voters subjected to disordered external
perturbations is affected by the fluctuations (variance and correlations) associated with the
external influence disorder. Our contributions can be summarized as follows. We show that random
inhomogeneities in the external perturbations tend to increase the critical values of the expected
perturbations compared with the no disorder counterpart. The size of the shift in the critical
behavior essentially depends on the fluctuations (variance and correlations) associated with the
multivariate distribution characterizing the strength of the external influence acting on voters.
The external perturbation disorder also has the surprising effect of amplifying the expected support
of an already biased opinion. The spontaneous emergence of the disorder-to-order phase transition is
marked by an increase in the variance of the equilibrium distribution, whose value is directly
related to the total fluctuation of the external influence. Our exact results for fully connected
networks also apply for quenched disordered networks, including random, regular lattice, scale-free,
and small-world networks.

% % % % % % % % % % % % % % % % % % % % % % % % % % % % % % % % % % %
% % % % % % % % % % % % % % % % % % % % % % % % % % % % % % % % % % %

\section{Model}
\label{model}

\subsection{Formulation}

Consider a network with $N$ voters endowed by two opinion states, denoted by 0 and 1. The external
influence disorder is represented by a nonnegative random vector ${\cal N} = (N_0,N_1)^T$  with
probability density function (pdf) $f(N_0,N_1)$, where $N_0$ and $N_1$ denote the strength of the
external influence towards opinions 0 and 1, respectively. At each time step, a node is randomly
selected. Let $k_0$ and $k_1$ denote the number of its nearest neighbors with opinions 0 and 1,
respectively. Then, the following events can occur: 1) with probability $p$, the state of the node
remains unchanged, 2) with probability $1-p$, the external influence vector ${\cal N} = (N_0,N_1)^T$
is randomly generated from the pdf $f(N_0,N_1)$, and is observed by the node. Then, the state of the
node becomes 0 with probability that is proportional to $N_0+k_0$, and becomes 1 with probability
that is proportional to $N_1+k_1$. Here we assume annealed disorder where the characteristic time
scale of changes in the external influence is comparable with the time scale of opinion dynamics.
When the external biases are zero, the above rule corresponds to a stochastic majority-vote rule
where a node randomly copies the state of one of its connected neighbors.

We initially assume that all voters can communicate with each other; so that the network of contacts
is a fully connected network (other topologies are considered later in the paper). In this case, the
behavior of the network can be solved exactly as follows. The nodes are indistinguishable and the
state of the network is fully specified by the number of nodes with internal state 1. Therefore,
there are only $N+1$ distinguishable global states, which we denote $S_k$,  $k=0,1,\dots,N$. The
state $S_k$ has $k$ nodes in state 1 and $N-k$ nodes in state 0. If $P_t(k)$ is the probability of
finding the network in state $S_k$ at time $t$, then $P_{t+1}(k)$ can depend only on $P_t(k)$,
$P_{t+1}(k)$ and $P_{t-1}(k)$. The dynamics is described by the equation

\begin{equation}
P_{t+1}(k) = \iint  P_{t+1}(k|N_0,N_1) f(N_0,N_1) dN_0 \, dN_1
\label{eq1}
\end{equation}
where the conditional probability $P_{t+1}(k|N_0,N_1)$ is given as follows:
\begin{equation}
\begin{array}{l}
P_{t+1}(k|N_0,N_1) = P_t(k)\Big\{ p +  \\ \\
 \displaystyle{	\left. \frac{(1-p)}
	{N(N+N_0+N_1-1)} \left[ k(k+N_1-1) +
	(N-k)(N+N_0-k-1) \right] \right\} +} \\ \\
P_t(k-1) \displaystyle{\frac{(1-p)}{N(N+N_0+N_1-1)}
	(k+N_1-1)(N-k+1)} \, + \\ \\
P_t(k+1) \displaystyle{\frac{(1-p)}{N(N+N_0+N_1-1)}
	(k+1)(N+N_0-k-1)} \;.
\end{array}
\label{eq2}
\end{equation}
The term inside the first brackets gives the probability that the state $S_k$ does not change in
that time step and is divided into two contributions: the probability $p$ that the node does not
change plus the probability $1-p$ that the node does change but copies another node in the same
state. In the latter case, the state of the node is 1 with probability $k/N$, and it may copy a
different node in the same state with probability $(k-1+N_1)/(N+N_0+N_1-1)$. Also, if the state of
the selected node is 0, which has probability $(N-k)/N$, it may copy another node in state 0 with
probability $(N-k-1+N_0)/(N+N_0+N_1-1)$. The other terms are obtained similarly.

Eq.(\ref{eq2}) can be rewritten as
\begin{equation}
\begin{array}{l}
P_{t+1}(k|N_0,N_1) =  P_t(k) \\ \\
	 \displaystyle{ -\frac{1-p}{N(N+N_0+N_1-1)} }\left\{ P_t(k) [2k(N-k)+N_1(N-k) +N_0 k ] \right. \\ \\    
     - P_t(k-1)  (k+N_1-1)(N-k+1) \\ \\
    \left. - P_t(k+1)	(k+1)(N+N_0-k-1)  \right\} \;.
\end{array}
\label{eq3}
\end{equation}

Exact results for the special no disorder case where $f(N_0,N_1) = \delta(N_0 - \bar{N}_0)
\delta(N_1 - \bar{N}_1)$ were presented in \cite{chinellato2007a, chinellato2015dynamical}. This
special case assumes that the external perturbations $N_0$ and $N_1$ are deterministic and fixed for
all times and nodes at their respective values of $\bar{N}_0$ and $\bar{N}_1$. Here we extend these
results to disordered opinion dynamics represented by any pdf $f(N_0,N_1)$ of the external
perturbations, and show how this external influence disorder affects the critical behavior of the
network compared with opinion dynamics without disorder. In this context, we also demonstrate an
interesting relationship between the moments of the bivariate distribution $f(N_0,N_1)$  and the
moments of the equilibrium distribution of the network.

\subsection{Analytical Results}

To address the general case, we notice that the integration in Eq \ref{eq1} leads to the following three integrals:
\begin{equation}
\frac{1}{n_T} \equiv \int \frac{f(N_0,N_1)}{N+N_0+N_1-1} dN_0 dN_1,
\label{eq4}
\end{equation}
\begin{equation}
\frac{n_0}{n_T} \equiv \int \frac{N_0 f(N_0,N_1)}{N+N_0+N_1-1} dN_0 dN_1
\label{eq5}
\end{equation}
and
\begin{equation}
\frac{n_1}{n_T} \equiv \int \frac{N_1 f(N_0,N_1)}{N+N_0+N_1-1} dN_0 dN_1.
\label{eq6}
\end{equation}
By plugging the definitions (\ref{eq4}-\ref{eq6}) into Eq. (\ref{eq1}), we obtain
\begin{equation}
\begin{array}{l}
P_{t+1}(k) = P_t(k) -  \displaystyle{\frac{(1-p)}{N n_T} } \Bigg(
P_t(k)[ 2k(N-k) + n_1(N-k) + n_0 k ]  \\ \\
 - P_t(k-1) 	(k+n_1-1)(N-k+1) \, - 
P_t(k+1) 	(k+1)(N+n_0-k-1)  \Bigg) \;.
\end{array}
\label{eq7}
\end{equation}

By comparing Eq.(\ref{eq7}) with Eq.(\ref{eq1}), the ``averaged'' parameters $n_0$ and $n_1$ can thus
be interpreted as the effective strengths of the external influence. Evaluating then the “averaged”
values $n_0$ and $n_1$ over the external influence pdf $f(N_0,N_1)$, the annealed disordered voter
model is mapped exactly to an effective voter model without disorder (i.e., where the external
influence strengths are fixed for all voters).

The probabilities $P_t(k)$ define a vector of $N+1$ components $\mathbf{P}_t$. In terms of $\mathbf{P}_t$, the master Eq.(\ref{eq7}) can be described by the equation
\begin{equation}
{\bf P}_{t+1} = {\bf U} {\bf P}_t \equiv 
\left( {\bf 1} - \frac{(1-p)}{N n_T} {\bf A}\right) {\bf P}_t 
\label{eq8}
\end{equation}
where the time evolution matrix $\bf{U}$, and also the auxiliary matrix $\bf{A}$, is tridiagonal. The non-zero elements of $\bf{A}$ are independent of $p$ and are given by
\begin{equation}
\begin{array}{l}
A_{k,k} = 2k(N-k) + n_1(N-k) + n_0 k \\
A_{k,k+1} = -(k+1)(N+n_0-k-1) \\
A_{k,k-1} = -(N-k+1)(n_1+k-1).
\end{array}
\label{eq9}
\end{equation}

The transition probability from state $S_M$ to $S_L$ after a time $t$ can be written as
\begin{equation}
P(L,t;M,0) = \sum_{r=0}^{N} b_{rM} \, a_{rL} \, \lambda_r^t
\label{eq10}
\end{equation}
where $a_{rL} $ and $b_{rM} $  are the components of the right and left $r$-th eigenvectors of the
evolution matrix, $\bf{a}_r$ and $\bf{b}_r$. Thus, the dynamical problem has been reduced to
finding the right and left eigenvectors and eigenvalues of the time evolution matrix $\bf{U}$.

The eigenvalues $\lambda_r$ of $\bf{U}$ are given by
\begin{equation}
\lambda_r = 1 - \frac{1-p}{N n_T} \, r (r-1+n_0+n_1) \qquad r=0,1,\dots, N
\label{eq11}
\end{equation}
and satisfy $0 \leq p \leq \lambda_r \leq 1$. The equation for $P(L,t;M,0) $ shows that the asymptotic
behavior of the network is determined only by the right and left eigenvectors with unit eigenvalue,
i.e., by the eigenvector corresponding to $\lambda_0 = 1$. The coefficients of the corresponding
(unnormalized) left eigenvector are simply $b_{0r} = 1$. The coefficients $a_{0r}$ of the right
eigenvector are obtained using a generating function technique and an associated nonlinear second
order differential equation \cite{chinellato2007a, chinellato2015dynamical}. The coefficients of the right eigenvector are then given by the
Taylor expansion of the hypergeometric function $F(-N,n_1,1-N-n_0,x) \equiv \sum_k a_{0k} \, x^k$.
After normalization, these coefficients give the stationary distribution of finding the network in
state $S_k$
\begin{equation}
\rho(k) = \frac{{{n_1+k-1}\choose{k}}{{N+n_0-k-1}\choose{N-k}}}  {{{N+n_0+n_1-1}\choose{N-k}}}
\label{eq12}
\end{equation}
This is the probability of finding the network with $k$ nodes in state 1 at equilibrium, and it is
independent of the initial state. The other eigenvectors, corresponding to $\lambda \neq 1$, can
also be calculated, and are also related to hypergeometric functions (4, 10). Although these
eigenvectors provide a complete description of the dynamics of the network (see Eq.(\ref{eq10})),
they are not particularly illuminating as we are interested in understanding the asymptotic behavior
of the system ($\lambda_0=1$).

In this paper, we are interested in the distributional properties of the fraction of nodes in state
1 -- that is, the vote-share $\nu = k/N$ -- rather than their number. The mean and variance of $\nu$
can be computed from Eq.(\ref{eq12}) as follows
\begin{equation}
\mu_{\nu} = \frac{n_1}{n_0+n_1}
\label{eq13}
\end{equation}
\begin{equation}
\sigma_{\nu}^2 = \frac{\mu_{\nu}(1-\mu_{\nu})}{N} \left( \frac{N}{n_0+n_1+1} + 
	\frac{n_0+n_1}{n_0+n_1+1} \right)
\label{eq14}
\end{equation}
The variance of vote-shares in Eq.(\ref{eq14}) has an appealing interpretation. When peer influences
(via social imitation) are very weak compared to “averaged” external forces ($n_0, n_1 \rightarrow
\infty$), the variance of vote-shares becomes $\sigma_{\nu}^2  =  \mu_{\nu}(1-\mu_{\nu})/N$. This is
the variance of vote-shares that one would expect if all nodes are solely influenced, each with
probability $\mu_{\nu}$, by the external biases, independent of the voting choices of other nodes.
The second term on the right side of Eq.(\ref{eq14}), which is a decreasing nonlinear function of
the external influence parameters, represents the effect of social imitation and peer influence
within the network. It is important to note that $\mu_{\nu}$ and $\sigma_{\nu}^2$ (as well as higher
moments) of the stationary vote-share distribution of the network both depend on the moments of the
bivariate distribution $f(N_0,N_1)$ governing the external influence strengths. A key result of our
paper is a characterization of this relationship, both analytically and numerically.

\subsection{Model behavior}
The stationary distribution 
$\rho(\nu)$ obtained from Eq \ref{eq12} has different shapes depending on the values of the effective parameters $n_0$ and $n_1$ defined in Eqs \ref{eq4}-\ref{eq6}. Similar to opinion dynamics without disorder \cite{chinellato2007a, chinellato2015dynamical}, as we move around in the 
$(n_0, n_1)$-parameter space, we observe different types of behavior, which is characteristic of a disorder-to-order critical phase transition.

For $n_0, n_1>1$ we obtain skewed unimodal distributions with peak at $n_1/(n_0+n_1)$ corresponding to the fraction of voters in the network that voted for opinion 1. 
If $n_1>n_0$ the majority of votes go to opinion 1, and if $n_0>n_1$ the majority of votes go to opinion 0. For $n_1>1$, $n_0<1$ or $n_1<1$, $n_0>1$ we obtain unimodal distributions with peaks at all nodes 1 or all nodes 0, respectively. 
For $n_0, n_1\gg 1$, $\rho(\nu)$ resembles a Gaussian distribution, and if $n_0 = n_1$ half the voters vote for opinion 0 and half the voters vote for opinion 1, similar to a magnetic material at high temperatures.

For $n_0, n_1<1$ -- the bistable ordered region -- we obtain bimodal distributions in which either of the two network phases can exist, similar to the magnetization state in the Ising model below the critical temperature. For $n_0=n_1\ll 1$, the distribution peaks at all nodes 0 or all nodes 1, similar to a magnetized state at low temperatures.

For $n_0 = n_1 = 1$ -- the \textit{critical value} of this model -- we obtain $\rho(k) = 1/(N+1)$ for all values of $N$. In this case, all states $S_k$ are equally likely and the system executes a random walk through the state space. In the limit $N\rightarrow\infty$, $n_0=n_1=1$ marks the critical transition between the disordered and ordered phases.

Finally, we note that for the symmetric case where the effective strengths of the external influence are equal $(n_0 = n_1 = n)$ the variance $\sigma_{\nu}^2$ of the stationary distribution $\rho(\nu)$ is a monotonically decreasing function of the effective strengths $\partial \sigma_{\nu}^2/\partial n_0<0$.
Therefore, the transition from the disordered unimodal phase  $(n_0 > 1)$ to the ordered bimodal phase $(n_0 < 1)$
    is associated with an \textit{increase} in the variability of the stationary distribution.

\section{Effect of disorder on critical behavior}

As shown above, the effective parameters $n_0$ and $n_1$ govern the critical behavior of the network. Moreover, the effective parameters, as defined in Eqs \ref{eq4}-\ref{eq6}, clearly depend on the moments of the external influence disorder
represented by the bivariate distribution $f(N_0, N_1)$. It is therefore interesting to study the effect of the fluctuations (variance and correlations) associated with the external influence disorder on the critical behavior.

To carry out the analysis, we apply a variety of approximations that become more exact in the thermodynamic limit of large number of nodes $N$. Let $\bar{N_0}=\av{N_0}$ and $\bar{N_1}=\av{N_1}$ be the mean values of the external influence strengths, and let $g(N_0, N_1)$ be any smooth function of the random vector $\mathbf{\mathcal{N}}=(N_0,N_1)^T$. We expand $g(N_0, N_1)$  up to second order to obtain

\begin{equation}
\begin{array}{ll}
\av{g(N_0, N_1)} & \equiv \int \int g(N_0, N_1)f(N_0, N_1)dN_0 dN_1 \\ \\
	& \approx g(\bar{N_0}, \bar{N_1}) + \frac{1}{2}\frac{\partial^2 g(\bar{N_0}, \bar{N_1})}{\partial N_0^2}\sigma_0^2 + \frac{\partial^2 g(\bar{N_0}, \bar{N_1})}{\partial N_0\partial N_1}\mbox{cov}_{01} +\frac{1}{2}\frac{\partial^2 g(\bar{N_0}, \bar{N_1})}{\partial N_1^2}\sigma_1^2 
\end{array}
	\label{eq15}
\end{equation}
where 
$\sigma_0^2\equiv \mbox{Var}(N_0)$, $\sigma_1^2\equiv \mbox{Var}(N_1)$, and $\mbox{cov}_{01}=\mbox{cov}(N_0, N_1)$. By applying Eq \ref{eq15} to the right hand side of Eqs \ref{eq4}-\ref{eq6}, we obtain

\begin{equation}
	\frac{1}{n_{T}}\approx\frac{1}{\bar{n_T}}\lrc{1+\frac{\sigma_0^2+2\mbox{cov}_{01}+\sigma_1^2}{\bar{n_T}^2}}
	\label{eq16}
\end{equation}

\begin{equation}
	\frac{n_0}{n_{T}}\approx\frac{1}{\bar{n_T}}\lrc{\bar{N_0}+\frac{-(N+\bar{N_1}-1)\sigma_0^2-(N+\bar{N_1}-\bar{N_0}-1)\mbox{cov}_{01}+\bar{N_0}\sigma_1^2}{\bar{n_T}^2}}
	\label{eq17}
\end{equation}

\begin{equation}
	\frac{n_1}{n_{T}}\approx\frac{1}{\bar{n_T}}\lrc{\bar{N_1}+\frac{-(N+\bar{N_0}-1)\sigma_1^2-(N-\bar{N_1}+\bar{N_0}-1)\mbox{cov}_{01}+\bar{N_1}\sigma_0^2}{\bar{n_T}^2}}
	\label{eq18}
\end{equation}
where $\bar{n_{T}}=N+\bar{N_0}+\bar{N_1}-1$. The factors $n_0$, $n_1$, and $n_T$ can be obtained explicitly from Eqs \ref{eq16}-\ref{eq18} as follows:

\begin{equation}
	n_0=\lrc{\frac{\bar{N_0}\bar{n_T}^2-\sigma_0^2(\bar{n_T}-\bar{N_0})-\mbox{cov}_{01}(\bar{n_T}-2\bar{N_0})+\bar{N_0}\sigma_1^2}{\bar{n_T}^2+\sigma_0^2+2\mbox{cov}_{01}+\sigma_1^2}}
	\label{eq19}
\end{equation}

\begin{equation}
	n_1=\lrc{\frac{\bar{N_1}\bar{n_T}^2-\sigma_1^2(\bar{n_T}-\bar{N_1})-\mbox{cov}_{01}(\bar{n_T}-2\bar{N_1})+\bar{N_1}\sigma_0^2}{\bar{n_T}^2+\sigma_0^2+2\mbox{cov}_{01}+\sigma_1^2}}
	\label{eq20}
\end{equation}

\begin{equation}
	n_{T}=N+n_0+n_1-1=\lrc{\frac{\bar{n_{T}}^3}{\bar{n_T}^2+\sigma_0^2+2\mbox{cov}_{01}+\sigma_1^2}}
	\label{eq21}
\end{equation}
For large number of nodes $N$, we expand Eqs \ref{eq19}-\ref{eq21} asymptotically in $N$ to obtain

\begin{equation}
	n_0=\bar{N_0}-\frac{\mbox{cov}_{01}+\sigma_0^2}{N}+\mathcal{O}\lrp{\frac{1}{N^2}}
	\label{eq22}	
\end{equation}

\begin{equation}
	n_1=\bar{N_1}-\frac{\mbox{cov}_{01}+\sigma_1^2}{N}+\mathcal{O}\lrp{\frac{1}{N^2}}
	\label{eq23}	
\end{equation}

\begin{equation}
	n_T=\bar{n_T}-\frac{\sigma_0^2+2\mbox{cov}_{01}+\sigma_1^2}{N}+\mathcal{O}\lrp{\frac{1}{N^2}}
	\label{eq24}	
\end{equation}

Eqs 
\ref{eq22}-\ref{eq24}
 relate the effective parameters $n_0$ and $n_1$ of the disordered
opinion dynamics model to the mean values $\bar{N_0}$ and $\bar{N_1}$ of the external influence strengths, and moreover show that for large networks with positively correlated external perturbations 
($\mbox{cov}_{01}>0$), $n_0<\bar{N_0}$ and $n_1<\bar{N_1}$. This implies that the critical values of the disordered opinion dynamics (i.e., $n_0=n_1=1$) can be
obtained even if the external influence strengths of the corresponding opinion dynamics without disorder satisfy $\bar{N_0}>1$ and $\bar{N_1}>1$. In other words, random inhomogeneities in the external perturbations tend to increase the critical values of the average perturbations compared with the critical values corresponding to opinion dynamics without disorder. The size of the shift in the critical behavior essentially depends on the fluctuations (variance and correlations) of the external influence disorder. Numerical simulations confirm this prediction (see Fig. \ref{fig:dists}).

We noted above that the spontaneous emergence of the disorder-to-order phase transition is often associated with an increase in the variability of the equilibrium distribution of voters' opinions. It would thus be interesting to study the effect of the fluctuations (variance and correlations) associated with the external influence disorder on the variability $\sigma_{\nu}^2$ of the equilibrium distribution. Eq \ref{eq12}  shows that the stationary distribution depends on the effective parameters $n_0$ and $n_1$. Moreover, as suggested by our previous discussion (Eqs \ref{eq19}-\ref{eq20}), the effective parameters essentially depend on the first and second moments of the external influence disorder represented by $f(N_0, N_1)$ (i.e., $\bar{N_0}, \bar{N_1}, \sigma_0^2, \sigma_1^2$ and $\mbox{cov}_{01}$). We therefore expect these moments to also approximate the moments of the stationary distribution (particularly the mean $\mu_{\nu}$ and variance $\sigma_{\nu}^2$). We characterize this relationship both analytically and numerically in the special symmetric case where $\mbox{var}(N_0)=\mbox{var}(N_1)\equiv\sigma^2$.

To carry out the analysis, we apply the approximations in Eqs 
\ref{eq19}-\ref{eq21} to the
mean $\mu_{\nu}$ and variance $\sigma_{\nu}^2$ shown in Eqs \ref{eq13}-\ref{eq14}. First, we observe from Eqs \ref{eq19}-\ref{eq21}
that, when $\sigma_0^2=\sigma_1^2\equiv\sigma^2$, both effective parameters $n_0$ and $n_1$ can be
expressed as a function of the total fluctuation, $\sigma_0^2+\mbox{cov}_{01}$, of the external
influence disorder (keeping $N$, $\bar{N_0}$, and $\bar{N_1}$ constant). But this also implies, from Eqs \ref{eq13}-\ref{eq14}, 
that both the first and second moments $\mu_{\nu}$ and $\sigma_{\nu}^2$ can be expressed
in terms of the total fluctuation $\sigma^2+\mbox{cov}_{01}$, rather than on each term individually. We can gain further insight into this relationship by asymptotically expanding Eqs \ref{eq13}-\ref{eq14} in $N$ to obtain

\begin{equation}
	\mu_{\nu}=\Phi_1 (\bar{N_0}, \bar{N_1}) + \Phi_2 (N, \bar{N_0}, \bar{N_1}) (\sigma^2+\mbox{cov}_{01})+\mathcal{O}\lrp{\frac{1}{N^2}}
	\label{eq25}
\end{equation}

\begin{equation}
	\sigma_{\nu}^2=\Psi_1 (N,\bar{N_0}, \bar{N_1}) + \Psi_2 (N, \bar{N_0}, \bar{N_1}) (\sigma^2+\mbox{cov}_{01})+\mathcal{O}\lrp{\frac{1}{N^2}}
	\label{eq26}
\end{equation}
where

\begin{equation}
	\Phi_1 (\bar{N_0}, \bar{N_1})=\frac{\bar{N_1}}{\bar{N_0}+\bar{N_1}}
	\label{eq27}
\end{equation}

\begin{equation}
	\Phi_2 (N, \bar{N_0}, \bar{N_1})=\frac{(\bar{N_1}-\bar{N_0})}{N(\bar{N_0}+\bar{N_1})^2}
	\label{eq28}
\end{equation}

\begin{equation}
	\Psi_1 (N, \bar{N_0}, \bar{N_1})=\frac{\bar{N_0}\bar{N_1}(\bar{N_0}+\bar{N_1}+N)}{N(\bar{N_0}+\bar{N_1})^2(\bar{N_0}+\bar{N_1}+1)}
	\label{eq29}
\end{equation}

\begin{equation}
	\Psi_2 (N, \bar{N_0}, \bar{N_1})=\frac{3\bar{N_0}^2\bar{N_1}+3\bar{N_0}\bar{N_1}^2+2\bar{N_0}\bar{N_1}-\bar{N_0}^2-\bar{N_1}^2-\bar{N_0}^3-\bar{N_1}^3}{N(\bar{N_0}+\bar{N_1})^3(\bar{N_0}+\bar{N_1}+1)^2}
	\label{eq30}
\end{equation}

We therefore expect, for large networks, a linear relationship between the total fluctuation $\mbox{cov}_{01}+\sigma^2$ and the first and second moments of the stationary distribution. Interestingly, according to Eqs \ref{eq25} and \ref{eq27}-\ref{eq28}, when there is an asymmetry between the average external influence biases (say $\bar{N_1}>\bar{N_0}$) the mean value $\mu_{\nu}$ of the stationary distribution is larger relative to the mean vote-share $\bar{N_1}/(\bar{N_0}+\bar{N_1})$ corresponding to opinion dynamics without disorder (i.e., when the external fluctuations are absent). In other words, if one opinion has an advantage over the other due to a larger expected external bias (e.g. $\bar{N_1}>\bar{N_0}$), this advantage will be amplified by the fluctuations of the external influence disorder.

If we further assume symmetry with respect to the mean strengths of the external influence, i.e. $\bar{N_0}=\bar{N_1}\equiv\bar{N}$, the following simplified expressions are obtained:

\begin{equation}
	\mu_{\nu}=\frac{1}{2}
	\label{eq31}
\end{equation}

\begin{equation}
	\sigma_{\nu}^2=\frac{1}{4N}\frac{N+2\bar{N}}{2\bar{N}+1}+\frac{1}{2N(2\bar{N}+1)^2}(\sigma^2+\mbox{cov}_{01})+\mathcal{O}\lrp{\frac{1}{N^2}}
	\label{eq32}
\end{equation}

It is interesting to note that in this case $\partial \sigma_{\nu}^2/\partial \sigma^2>0$ and $\partial \sigma_{\nu}^2/\partial \mbox{cov}_{01}>0$. Moreover, for positively correlated external perturbations ($\mbox{cov}_{01}>0$), $\partial \sigma_{\nu}^2/\partial N<0$. In other words, the variance $\sigma_{\nu}^2$ of the stationary distribution $\rho(\nu)$ is monotonically increasing in the fluctuations of the external influence disorder, and is monotonically decreasing in the mean strengths of the external influence. Therefore, following our previous discussion, as we decrease the mean strengths, and as we increase the fluctuations, the variability of the stationary distribution is increased, and the distribution becomes progressively flatter. This increased variability marks the transition from the disordered unimodal phase to the ordered bimodal phase. Numerical simulations confirm this prediction (see Fig. \ref{fig:dists}).

\section{Computational results} 

Here, we consider the case where the external influence vector 
$\mathbf{\mathcal{N}}=(N_0,N_1)^T$ is drawn from a bivariate lognormal distribution. The bivariate lognormal distribution is chosen for analytic tractability, although its behavior is also quite natural for representing the external influence disorder: the bivariate lognormal distribution is useful in modeling correlated multivariate heavy- tailed data, which appears frequently in the physical and social sciences \cite{limpert2001log, yue2000bivariate, kleiber2003statistical}. Using the bivariate lognormal distribution, we perform a variety of computational experiments, testing our analytical results.

More specifically, the external influence strengths 
$\mathbf{\mathcal{N}}=(N_0,N_1)^T$ 
are given by 
$\mathbf{\mathcal{N}}=e^{\mathbf{X}}$, 
where $\mathbf{X}=(X_0, X_1)^T$ 
is a bivariate normal distribution with mean 
$\mu=(\mu_1, \mu_2)^T$, 
covariance matrix
$$
\Sigma = 
	\begin{pmatrix}
		\mbox{var}(X_1) & \mbox{cov}(X_1, X_2)  \\
		\mbox{cov}(X_2, X_1) & \mbox{var}(X_2)
	\end{pmatrix}
	=
	\begin{pmatrix}
		\sigma_{X_1}^2 & \rho\sigma_{X_1}\sigma_{X_2}  \\
		\rho\sigma_{X_2}\sigma_{X_1} & \sigma_{X_2}^2
	\end{pmatrix}
$$
and correlation $\rho\equiv\mbox{cor}(X_1, X_2)=\frac{\mbox{cov}(X_1, X_2)}{\sigma_{X_1}\sigma_{X_2}}$. The mean and covariance matrix of the random vector
$\mathbf{\mathcal{N}}=e^\mathbf{X}$ are given as follows:

\begin{equation}
	\mathbf{\mathcal{\bar{N}}}=
	\begin{pmatrix}
		\bar{N_0} \\
		\bar{N_1}
	\end{pmatrix}
	=
	\begin{pmatrix}
		e^{\mu_1+\frac{\sigma_{X_1}^2}{2}} \\
		e^{\mu_2+\frac{\sigma_{X_2}^2}{2}}
	\end{pmatrix}	
\end{equation}

\begin{eqnarray}
	\mathbf{V} &=&
	\begin{pmatrix}
		\sigma_0^2 & \mbox{cov}_{01} \\
		\mbox{cov}_{10} & \sigma_1^2
	\end{pmatrix} \\ \nonumber
	&=&
	\begin{pmatrix}
		e^{2\mu_1+\sigma_{X_1}^2}(e^{\sigma_{X_1}^2}-1) & e^{\mu_1+\mu_2+\frac{1}{2}(\sigma_{X_1}^2+\sigma_{X_2}^2)}(e^{\mbox{cov}(X_1, X_2)}-1) \\
		e^{\mu_1+\mu_2+\frac{1}{2}(\sigma_{X_1}^2+\sigma_{X_2}^2)}(e^{\mbox{cov}(X_2, X_1)}-1) & e^{2\mu_2+\sigma_{X_2}^2}(e^{\sigma_{X_2}^2}-1)
	\end{pmatrix}
\end{eqnarray}

In the simulations below, we consider two cases: 
(1) $\sigma_0^2=\sigma_1^2\equiv\sigma^2$ and 
$\bar{N_0}=\bar{N_1}\equiv\bar{N}$, 
and
(2) $\sigma_0^2=\sigma_1^2\equiv\sigma^2$ and $\bar{N_0}\neq\bar{N_1}$.
For both cases, we study the effect of the total fluctuation 
$\sigma^2 + \mbox{cov}_{01}$ on the critical behavior of the model as well as on the first and second moments $\mu_{\nu}$ and $\sigma_{\nu}^2$ of the stationary distribution. In the appendix, we describe in detail the methods used to generate the bivariate lognormal distributions for the various simulations.

\begin{figure}[h!]
	\begin{center}
	    \includegraphics[width=0.95\columnwidth]{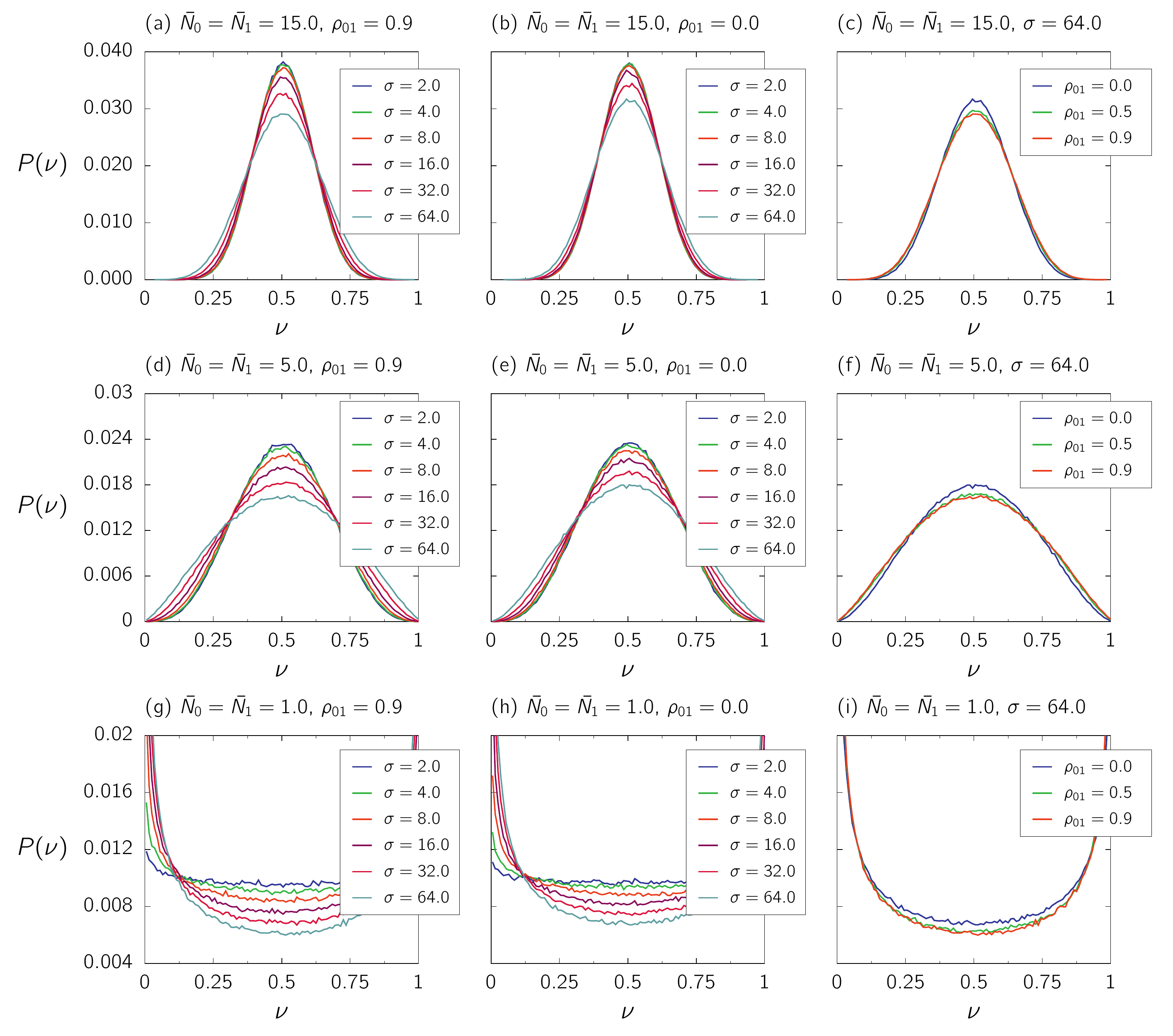}
	\end{center}
	\vspace{-0.75cm}
	\caption{\textbf{The effect of the external influence disorder on the equilibrium distribution and critical behavior of the network}. Here, different stationary vote-share distributions are generated by varying the parameters of the bivariate lognormal distribution: $\bar{N_0}=\bar{N_1}\equiv\bar{N}$, $\sigma_0^2=\sigma_1^2\equiv\sigma^2$ and $\rho_{01}$. The simulations are performed on fully connected networks with $N=100$ voters. The stationary distributions are determined by running each simulation for $15\times 10^3$ time steps, and then averaging over $10^6$ measurements collected at intervals of $2\times 10^3$ time steps. We see that the stationary distribution becomes progressively flatter for increasing values of $\sigma^2$ and $\mbox{cov}_{01}$, and for decreasing values of $\bar{N}$.} 
	\label{fig:dists}
\end{figure}

In Fig.~\ref{fig:dists} we show, for a fully connected network, the effect of the external influence
disorder on the equilibrium distribution and critical behavior of the network, by varying the values
of the mean strengths $\bar{N_0}=\bar{N_1}\equiv\bar{N}$ variabilities
$\sigma_0^2=\sigma_1^2\equiv\sigma^2$, and correlation $\rho_{01} \equiv \mbox{cov}_{01}/\sigma^2$.
The results confirm the theoretical findings. As we decrease the mean intensity of the external
influence, and increase the fluctuations (variance and correlations), the variability of the
stationary distribution is increased, and the distribution becomes progressively flatter. Moreover,
the increased variability of the stationary distribution is marked by a transition from disordered
unimodal distributions to ordered bimodal distributions. However, more interestingly, the bimodal
phase is already observed at $\bar{N_0}=\bar{N_1}=1$ -- the critical values that correspond to
opinion dynamics without disorder (i.e., $\sigma^2=0$ and $\rho_{01}=0$). In other words, for fixed
values of $\sigma^2$ and $\rho_{01}$, the critical values of the disordered opinion dynamics model
are obtained for $\bar{N_0}>1$ and $\bar{N_1}>1$. For these critical values, the equilibrium
distribution becomes uniform (i.e., $\rho(k) = 1/(N+1)$), marking the transition between the
disordered and ordered phases. The size of the shift in the critical behavior -- relative to the
critical values corresponding to opinion dynamics without disorder -- essentially depends on the
magnitude of the fluctuations (variance and correlations) of the external influence disorder. As
suggested by Fig.~\ref{fig:dists}, the larger the fluctuations, the larger the shift in the critical
behavior. What do these results mean? Recall that voters observe the external influence biases and
the voting of other agents. For large perturbations $n_0$ and $n_1$, the external influences extend
into the network, dwarfing the effect of peer interactions within the network. For small
perturbations, on the other hand, the effect of peer influence dominates the effect of external
influence. This latter case is the origin of increased variability, and is a manifestation of
self-organized, collective behavior of the network. Fig.~\ref{fig:dists} then tells us that the
self-organized, collective behavior of the network is driven not only by the mean external
influences (as in opinion dynamics without disorder) but also by the uncertainty and correlations
present in the external environment. The critical behavior observed in Fig.~\ref{fig:dists} is
therefore an example of fluctuation induced critical phase transition.

\begin{figure}[h!]
	    \includegraphics[width=0.95\columnwidth]{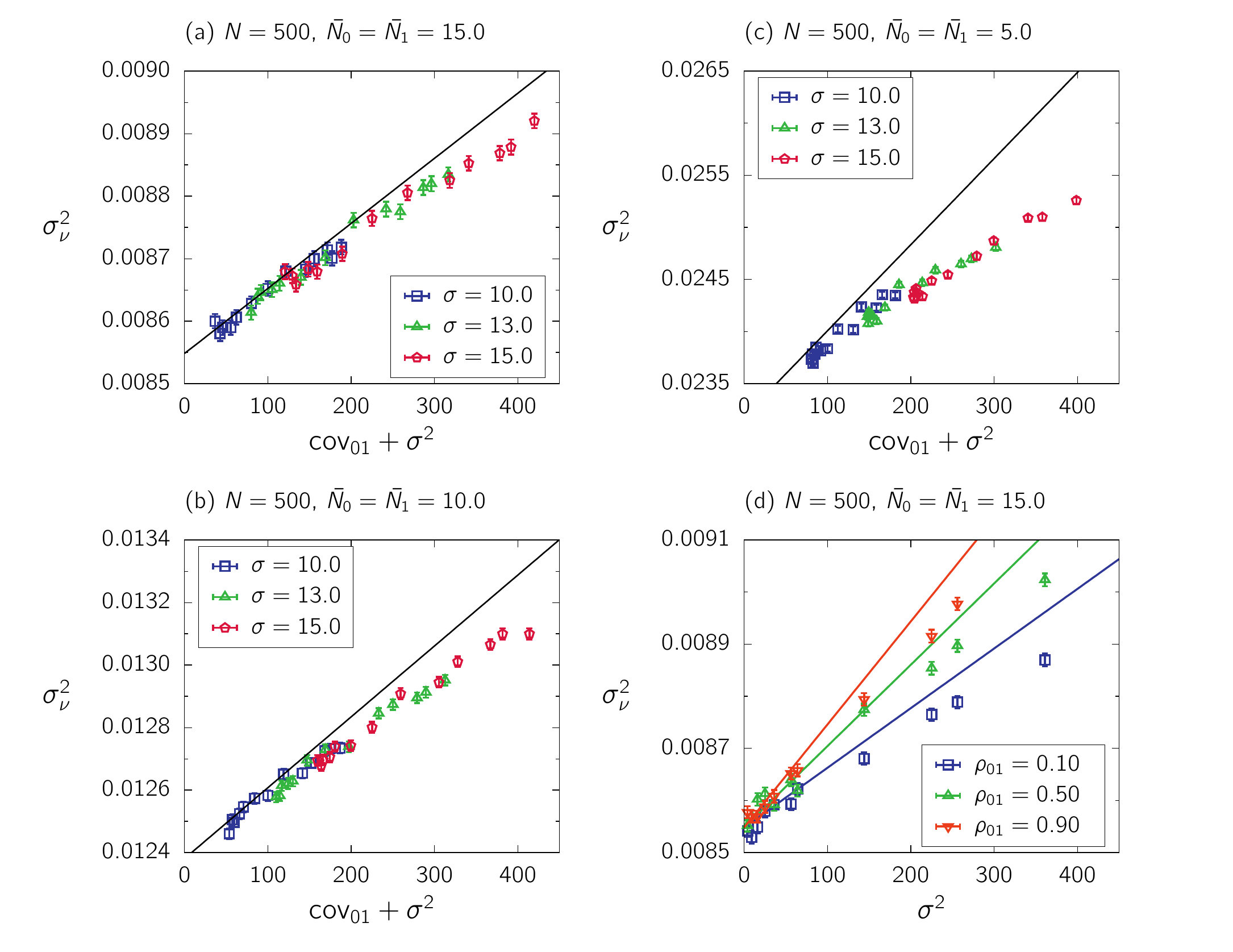}
	\vspace{-0.75cm}
	\caption{\textbf{The effect of total fluctuation on vote-share variability: the fully symmetric case.} The solid lines are analytical results (see Eq \ref{eq32}); the symbols are obtained from Monte Carlo simulations for fully connected networks with $N=500$ voters, assuming fully symmetric bivariate lognormal distributions $\bar{N_0}=\bar{N_1}\equiv\bar{N}$ and $\sigma_0^2=\sigma_1^2\equiv\sigma^2$. 
	The stationary distributions are determined by running each simulation for $10^5$ time steps, and then averaging over $10^6$ measurements collected at intervals of $10^4$ time steps. The error bars of $\sigma_{\nu}^2$ represent the bootstrap standard error \cite{efron1994introduction, newman1999monte}, calculated through $10^3$ resamplings of the original data. 
	a-c) The variance $\sigma_{\nu}^2$ of the stationary distribution as a function of $\sigma^2+\mbox{cov}_{01}$, for various $\bar{N}$ and $\sigma^2$. We note that $\sigma_{\nu}^2$ can be expressed as a function of the total fluctuation $\sigma^2+\mbox{cov}_{01}$, rather than each term separately. d) The variance $\sigma_{\nu}^2$ as a function of $\sigma^2$, for various values of $\rho_{01}$ (effectively, $\mbox{cov}_{01}$). We see that the variability $\sigma_{\nu}^2$ is increasing in both $\sigma^2$ and $\mbox{cov}_{01}$, in accordance to theory.
	} 
	\label{fig:sim_fc_nets}
\end{figure}

\begin{figure}[h!]
	\begin{center}
	    \includegraphics[width=0.95\columnwidth]{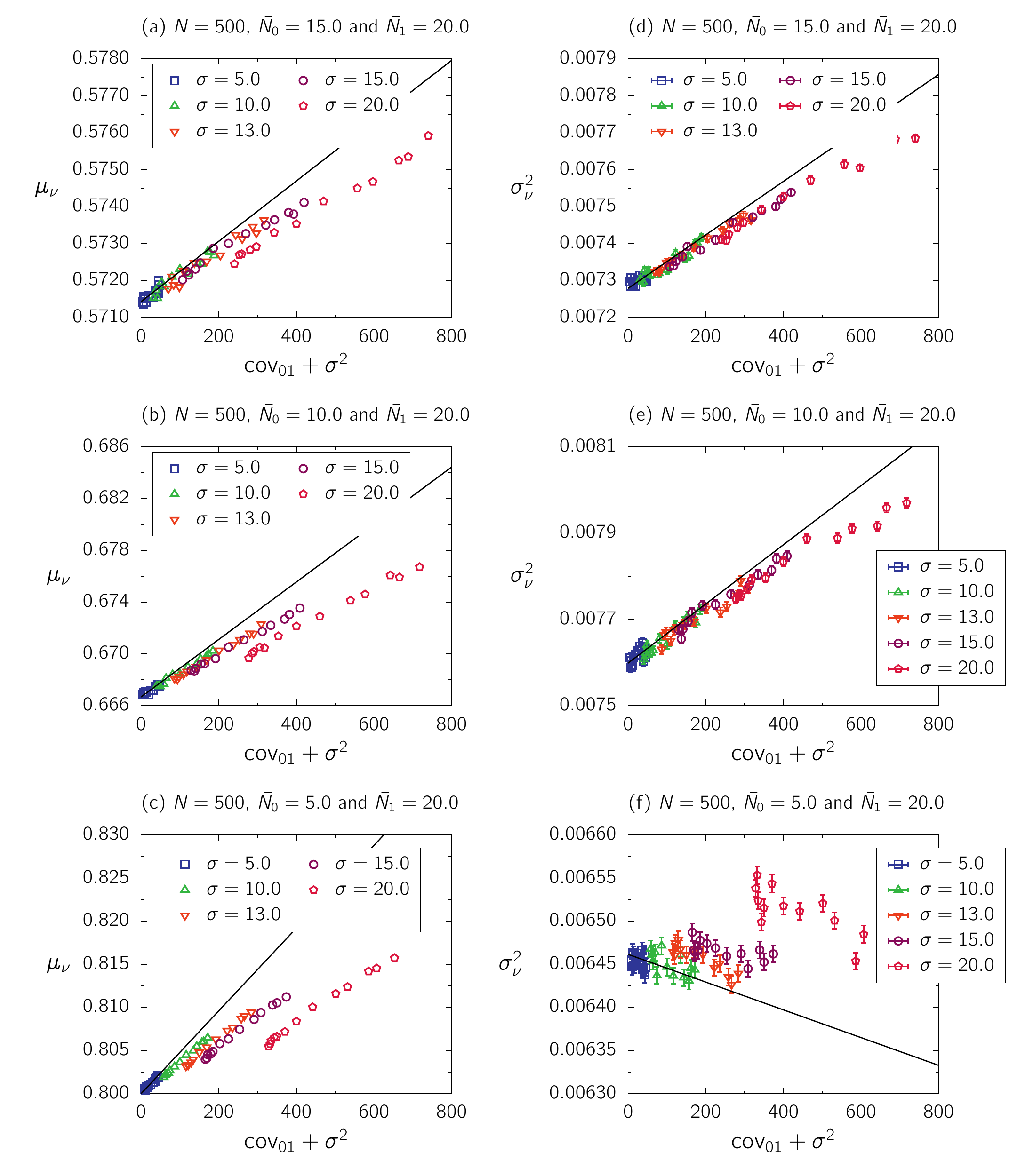}
	\end{center}
	\vspace{-0.75cm}
	\caption{
	\textbf{The effect of total fluctuation on vote-share variability: the semi-symmetric case.} 
	The solid lines are analytical results (see Eqs \ref{eq25}-\ref{eq26}); the symbols are obtained from Monte Carlo simulations for fully connected networks with $N=500$ voters, assuming semi-symmetric bivariate lognormal distributions ($\bar{N_0}\neq\bar{N_1}$ and $\sigma_0^2=\sigma_1^2\equiv\sigma^2$).
The stationary distributions and error bars of $\sigma_{\nu}^2$ are determined as described in Fig.~\ref{fig:sim_fc_nets}. 
a-c) The mean $\mu_{\nu}$ of the stationary distribution as a function of $\sigma^2+\mbox{cov}_{01}$, for various $\bar{N_0}$, $\bar{N_1}$ and $\sigma^2$. 
d-f) The variance $\sigma_{\nu}^2$ of the stationary distribution as a function of $\sigma^2+\mbox{cov}_{01}$, for various $\bar{N_0}$, $\bar{N_1}$ and $\sigma^2$. We note that both $\mu_2$ and $\sigma_{\nu}^2$ can be expressed as a function of the total fluctuation $\sigma^2+\mbox{cov}_{01}$, rather than each term separately.} 
	\label{fig:as_fc_nets}
\end{figure}

We noted in Fig.~\ref{fig:dists} that the disorder-to-order phase transition is associated with an
increase in the variability of the equilibrium distribution. As suggested earlier, this variability
(as well as the first moment) can be expressed directly as a function of the total fluctuation,
$\sigma^2+\mbox{cov}_{01}$, of the external influence disorder. The numerical simulations confirm
this dependence for a wide range of external influence fluctuations. The simulation results of the
fully symmetric case ($\sigma_0^2=\sigma_1^2\equiv\sigma^2$ and $\bar{N_0}=\bar{N_1}\equiv\bar{N}$ )
are shown in Fig.~\ref{fig:sim_fc_nets}, whereas the semi-symmetric case
($\sigma_0^2=\sigma_1^2\equiv\sigma^2$, $\bar{N_0}\neq\bar{N_1}$) is shown in
Fig.~\ref{fig:as_fc_nets}. As the figures show, the simulation results are in good agreement with
the analytical predictions of Eqs 25-26 and 31-32. More specifically, we find that the first and
second moments of the equilibrium distribution are dependent on the total fluctuation
$\sigma^2+\mbox{cov}_{01}$, rather than on each term individually. This dependence is seen to be
nearly linear in $\sigma^2+\mbox{cov}_{01}$ for a wide range of the total fluctuation. Finally, for
the fully symmetric case, the variability $\sigma_{\nu}^2$ is seen to be increasing with both
$\sigma^2$ and $\mbox{cov}_{01}$ (Fig.~\ref{fig:sim_fc_nets}d), and decreasing with the mean values
$\bar{N_0}=\bar{N_1}$ of the external influence strengths, in alignment with the analytical results
and Fig.~\ref{fig:dists}. We note that in the absence of fluctuations of the external influence
disorder ($\sigma^2=0$, $\mbox{cov}_{01} = 0$), the mean vote-share of the first opinion is given by
$\mu=\bar{N_1}/(\bar{N_0}+\bar{N_1})$. In this case, for $\bar{N_1}>\bar{N_0}$ as in
Fig.~\ref{fig:as_fc_nets}, the first opinion has an advantage (on average) over the other. However,
interestingly and unexpectedly, the introduction of disorder and fluctuations in the external
perturbations ($\sigma^2>0$, $\mbox{cov}_{01}>0$) leads to an increase in the mean vote-share
$\mu_{\nu}$ (Fig.~\ref{fig:as_fc_nets}a-c) relative to the mean vote-share
$\bar{N_1}/(\bar{N_0}+\bar{N_1})$ associated with opinion dynamics without disorder. This is
consistent with our theoretical prediction that the fluctuations in the external perturbations have
the effect of amplifying the expected support of an opinion that already has an advantage.

\begin{figure}[h!]
	\begin{center}
	    \includegraphics[width=0.95\columnwidth]{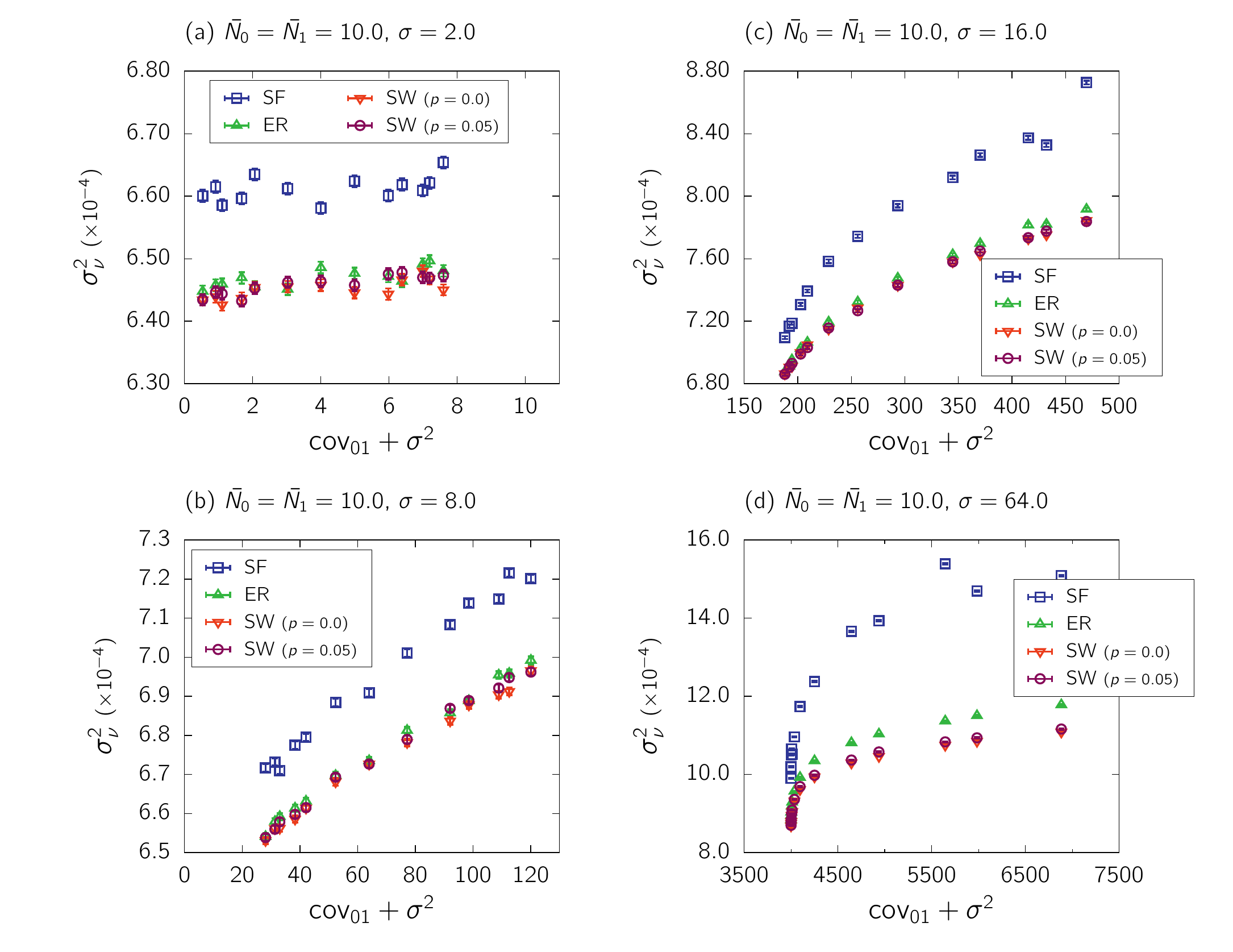}
	\end{center}
	\vspace{-0.75cm}
	\caption{
\textbf{The effect of total fluctuation on vote-share variability for networks with different topologies.} Results are obtained from Monte Carlo simulations for networks with $N=500$ voters and average connectivity $\av{k}=6$, assuming fully symmetric bivariate lognormal distributions ($\bar{N_0}=\bar{N_1}\equiv\bar{N}$ and $\sigma_0^2=\sigma_1^2\equiv\sigma^2$). 
Here we use scale-free (SF), random Erd\H{o}s-Rényi (ER), small-world (SW, rewiring probability $p=0.05$), and regular ring lattice (SW, rewiring probability $p=0$). The stationary distributions and error bars of $\sigma_{\nu}^2$ are determined as described in Fig. 2. 
The figure shows the variance $\sigma_{\nu}^2$  of the stationary distribution as a function of $\sigma^2+\mbox{cov}_{01}$, for various $\bar{N}$ and $\sigma^2$. 
We see that $\sigma_{\nu}^2$ can be expressed roughly as a linear function of the total fluctuation $\sigma^2+\mbox{cov}_{01}$, similar to fully connected networks.
} 
	\label{fig:sim_random_nets}
\end{figure}

Although the stationary distribution given by Eq \ref{eq12} is obtained assuming fully connected networks, we also consider a quenched disorder that arises from randomness in the topology of the network. Here we present simulation results obtained for different topologies with the same average connectivity, including random, regular lattice, scale-free, and small-world networks. Fig.~\ref{fig:sim_random_nets} shows the variance $\sigma_{\nu}^2$ of the stationary distribution as a function of the total fluctuation $\sigma^2 + \mbox{cov}_{01}$, for various $\bar{N}$ and $\sigma^2$. 
Interestingly, the results show that the analytical insight gained for fully connected networks applies qualitatively for other topologies as well: the variability $\sigma_{\nu}^2$ can be expressed as a function of the total fluctuation $\sigma^2 + \mbox{cov}_{01}$, and is roughly linear in $\sigma^2 + \mbox{cov}_{01}$. Figs~\ref{fig:sim_fc_nets} and \ref{fig:sim_random_nets} also tell us that for topologies that are not fully connected the variability of the stationary distribution is much smaller compared to fully connected networks with the \textit{same} number of voters and total fluctuation $\sigma^2 + \mbox{cov}_{01}$. This means that for topologies that are not fully connected the perturbations of the external environment extend more easily and rapidly into the network, weakening the peer influence effects. In other words, networks with smaller connectivity will tend to weaken the perturbations within the network, and amplify the effect of the external influence. Finally, Fig.~\ref{fig:sim_random_nets} shows that larger variabilities are observed for the scale-free network relative to the other topologies. Thus the scale-free topology seems to propagate the perturbations within the network more effectively relative to other networks. We conjecture that the critical behavior of opinion networks is affected not only by heterogeneities in the external environment, but also by the large connectivity fluctuations usually found in heterogeneous networks.

%%%%%%%%%%%%%%%%%%%%%%%%%%%%%%%%%%%%%%%%%%%%%%%%%%%%%%%%%
\section{Summary}
\label{conclusions}

In this paper, we analyzed an influence network of voters subjected to correlated disordered external perturbations. We showed that the random heterogeneities in the external perturbations affect the critical behavior of the network. The size of the shift in the critical behavior, relative to networks without disorder, essentially depends on the total fluctuations of the external influence disorder. We demonstrated that the model exhibits a critical phase transition, which is marked by an increase in the variance of the equilibrium distribution. We found analytically that this variance is directly related to the total fluctuation of the external influence. We extended our analysis by considering a fat-tailed multivariate lognormal disorder, and presented numerical simulations that confirmed our analytical results. Simulations for different network topologies showed that similar results apply to other networks as well.

Our work can be extended in several ways. In this paper we considered the case of annealed disorder of the external perturbations. It would be interesting to extend the analysis developed here to the quenched disorder case in which the characteristic time scale of changes in the external influence is much longer than the characteristic time scale of the voter opinion fluctuations. In this case, the values of the random variables vary from one voter to another, but remain constant in time. In our simulations, we also considered a quenched disorder that arises from randomness in the topology of the network. Here, we assumed that the underlying network structure is fixed, and therefore the network does not evolve with time. However, real networks are often dynamic and evolve rapidly with time\cite{braha2006centrality, hill2010dynamic}, and the assumption of quenched disorder would not be valid if the characteristic time scale for changes in the network is comparable with the time scale of opinion dynamics. In this case, the disorder should be considered to be annealed. Finally, as suggested in Fig.~\ref{fig:sim_random_nets}, we hope to be able to quantify the effect of heterogeneous networks with large connectivity fluctuations on the behavior of opinion networks.

\begin{acknowledgements}
MFR is supported by CNPq (grant $\#$152885/2016-1). MAMA was partially
supported by Fapesp (grant $\#$2016/05460-3) and CNPq (grant $\#$302049/2015-9).
\end{acknowledgements}

\section*{Appendix: Generating the lognormal distributions}
\label{sec:appendix}

In this appendix, we describe the methods used to generate the bivariate lognormal distributions for the various simulations.

Consider a semi-symmetric bivariate lognormal distributions ($\bar{N_0}\neq\bar{N_1}$ and $\sigma_0^2=\sigma_1^2\equiv\sigma^2$) as used in Fig. 3. In this case, we fix the values of $\bar{N_0}, \bar{N_1}, \sigma$ and $\rho_{01}$,  and determine the values of  $\mu_1$, $\mu_2$, $\sigma_{X_1}$, $\sigma_{X_2}$ and $\rho$ by solving Eq 33 and 34. From these Eqs we can extract the values of $\sigma_{X_1}$ and $\mu_1$: 
 \begin{equation}
 	\sigma_{X_1}^2=\ln\lrp{\frac{\sigma^2}{\bar{N_0}^2}+1}
 \end{equation}
and 
 \begin{equation}
	\mu_1=\ln(\bar{N_0})-\frac{\sigma_{X_1}^2}{2}.
 \end{equation}
The values of $\sigma_{X_2}$ and $\mu_2$ can be obtained  similarly. The covariance $\mbox{cov}(X_1, X_2)$ is extracted from Eq 34:
 \begin{equation}
 	\mbox{cov}(X_1, X_2)=\ln\lrp{\frac{\rho_{01}\sigma^2}{\bar{N_0}\bar{N_1}}+1}.
 \end{equation}

 Finally, the expression for $\rho$ is obtained from 
 $\rho=\mbox{cov}(X_1, X_2)/(\sigma_{X_1}\sigma_{X_2})$.
 Having determined the values of $\mu_1$, $\mu_2$, $\sigma_{X_1}$, $\sigma_{X_2}$ and $\rho$ we can generate a bivariate normal variable 
 $(X_1, X_2)^{T}$, which is then used to generate the external influence vector $\mathcal{N}=e^{\mathbf{X}}$.
 The  fully symmetric bivariate lognormal distributions
 as the ones used in Figs. 1, 2 and 4 can be obtained by setting $\bar{N_0}=\bar{N_1}\equiv\bar{N}$.

\bibliographystyle{unsrt}

\end{document}